\documentclass[11pt,a4paper]{article} 
\usepackage{jheppub}

\usepackage{slashed}
\usepackage{youngtab}

\newcommand{\la}{\langle}
\newcommand{\ra}{\rangle}

\newcommand{\beq}{\begin{equation}}
\newcommand{\eeq}{\end{equation}}

\newcommand{\beqa}{\begin{eqnarray}}
\newcommand{\eeqa}{\end{eqnarray}}

\newcommand{\by}{\begin{eqnarray}}
\newcommand{\ey}{\end{eqnarray}}
\newcommand{\al}{\alpha}
\newcommand{\be}{\beta}
\newcommand{\ga}{\gamma}

\newcommand{\de}{\delta}
\newcommand{\ep}{\epsilon}
\newcommand{\om}{\omega}
\newcommand{\si}{\sigma}
\newcommand{\half}{\frac{1}{2}}
\newcommand{\quart}{\frac{1}{4}}
\newcommand{\thalf}{\tfrac{1}{2}}

\newcommand{\pd}{\partial}
\newcommand{\lam}{\lambda}

\newcommand{\tM}{\tilde M}
\newcommand{\tP}{\tilde P}
\newcommand{\tK}{\tilde K}
\newcommand{\tD}{\tilde D}
\newcommand{\te}{\tilde e}
\newcommand{\tf}{\tilde f}
\newcommand{\tom}{\tilde \om}
\newcommand{\tb}{\tilde b}
\newcommand{\tLam}{\tilde \Lambda}

\newcommand\fverb{\setbox\fverbbox=\hbox\bgroup\verb}
\newcommand\fverbdo{\egroup\medskip\noindent%
            \fbox{\unhbox\fverbbox}\ }
\newcommand\fverbit{\egroup\item[\fbox{\unhbox\fverbbox}]}
\newbox\fverbbox

\newcommand{\nablaslash}{\not{\hbox{\kern-3pt $\nabla$}}}

\title{On the conformal higher spin unfolded equation for a three-dimensional self-interacting scalar field}
\author{Bengt E.W.~Nilsson%\\
\\
Fundamental Physics\\
Chalmers University of Technology\\
SE-412 96 G\"oteborg, Sweden\\

{\tt {\footnotesize  tfebn@chalmers.se}}}

\abstract{ We propose  field equations for the  conformal higher spin system in three dimensions coupled to  a conformal scalar field 
with a sixth order potential. Both the higher spin equation and the unfolded equation for the scalar field have source terms
and  are based on a conformal higher spin algebra which we treat as an expansion in multi-commutators. Explicit expressions for 
the source terms are suggested and subjected to some simple tests. We also discuss a cascading relation between the Chern-Simons 
action for the higher spin gauge theory and an action containing a term for each spin that generalizes the spin 2 Chern-Simons action 
in terms of the spin connection expressed in terms of the frame field. This cascading property 
is demonstrated in the free theory for spin 3 but should work also in the complete higher spin theory.}


\toccontinuoustrue
\begin{document}
\maketitle

%\setcounter{page}{2}

%%%%%%%%%%%%%%%%%%%%%%%%%%%%%%%%%%%%%%%%%%%%%%%%%%%%%%%%%%%%

\section{Introduction}

%%%%%%%%%%%%%%%%%%%%%%%%%%%%%%%%%%%%%%%%%%%%%%%%%%%%%%%%%%%%

The purpose of this  paper is to propose a set of  field equations in three dimensions that describe a fully interacting conformal system
consisting of a scalar field and the higher spin theory generated by  the $SO(3,2)$ higher spin algebra. We will follow the work
 \cite{Nilsson:2013tva} where this approach to the problem was discussed in some detail and some  results were found indicating that this 
 may be worth pursuing further.
 
After giving the proposed equations in section 2 we will first explain the notation and content of them  and then present some of the arguments 
leading to their particular form together with  some explicit checks  that will give some support 
for  this proposal. 
The higher spin part of the theory has its origin in a gauged $SO(3,2)$ Chern-Simons theory which can be reformulated as a generalization 
to all higher spins of the standard spin 2 Chern-Simons theory for the  spin connection. This will be elaborated upon
in section 3 where a cascading trick is used to relate the two different Chern-Simons formulations of the higher spin theory. Some additional comments are collected in the Conclusions.

Thus our main goal will be to present two higher spin equations, one field strength and one unfolded equation, and to show that the 
following spin 0 (Klein-Gordon) and spin 2 (Cotton)  equations can be reproduced:
\beq
\label{kleingordon}
\Box\phi-\frac{1}{8}R\phi-\frac{27g^3}{32 \cdot 32}\phi^5=0,
\eeq
and
\beqa
\label{cottoneq}
&&C_{\mu\nu}-\frac{g}{16}(R_{\mu\nu}-\half g_{\mu\nu}R)\phi^2-\frac{g}{2}(\pd_{\mu}\phi\pd_{\nu}\phi-\half g_{\mu\nu}\pd^{\rho}\phi\pd_{\rho}\phi)\cr
&&-\frac{g}{8}(\phi\Box\phi+\pd^{\rho}\phi\pd_{\rho}\phi)g_{\mu\nu}+\frac{g}{8}(\phi D_{\mu}\pd_{\nu}\phi+\pd_{\mu}\phi\pd_{\nu}\phi)+\frac{9g^3}{2\cdot 32 \cdot 32}\phi^6=0.
\eeqa
Note, however,  that these equations are taken from the topological gauging of three dimensional $CFT$s with eight supersymmetries \cite{Gran:2008qx, Gran:2012mg} where all coupling constants are determined in terms 
of the gravitational one $g$. This needs not to be the case in non-supersymmetric theories like the ones we deal with here. 

Once the higher spin equations are presented we can
discuss their consequences for the field equations for spin 3 and above. Only a few such comments will be given here while a more extensive survey will be left for a future publication. We may note already at this point that the spin 2 equation above will be augmented by new terms with more than two derivatives 
of the scalar $\phi(x)$ provided higher spin frame fields also appear. This is true also for the equations of  spin 3 etc and follows directly  from the fact that $\phi$ has conformal dimension $L^{-\half}$ and that the number of derivatives in the spin $s$ equation is $2s-1$ which implies that 
the spin $s$ frame field itself is of dimension $L^{s-2}$. Also the Klein-Gordon equation will contain terms with higher spin fields and more than two derivatives  on the scalar. 

We will find it convenient to write the Cotton equation (\ref{cottoneq}) in irreps of $SO(1,2)$. The point is  that the   trace is exactly the Klein-Gordon 
equation which means that  the rest of the Cotton equation  is in the irrep ${\bf 5}$ and reads
\beq
\label{spintwocotton}
C_{\mu\nu}-\frac{g}{16}(\phi^2R_{\mu\nu}-2\phi D_{\mu}\pd_{\nu}\phi+6\pd_{\mu}\phi\pd_{\nu}\phi)|_{\bf 5}=0,
\eeq
where we recall that the Cotton tensor is already in this irrep. The purpose of this paper is to suggest two higher spin field equations containing component 
equations for all spins $\geq 2$
coupled to a scalar field $\phi$ with $\phi^6$ potential  and which in particular reproduce both the above Klein-Gordon and spin 2 Cotton equations. 
Already in  
\cite{Nilsson:2013tva} where this approach was discussed, but without source terms in either equation, it was shown that  the correct curvature scalar term does arise 
in the Klein-Gordon equation and in addition also a spin 3 contribution\footnote{There are probably also spin 3 terms  with one or more derivatives on the scalar field.}
\beq
\Box\phi-\frac{1}{8}R\phi+\tf\phi=0,
\eeq
where the spin 3 term contains the trace $\tf:=e^{\mu}{}_a\tf_{\mu}{}^a(1,3)$.  The field  $\tf_{\mu}{}^a(1,3)$, which is an expression containing three derivatives on the spin 3 frame field $e_{\mu}{}^{ab}$ 
is  discussed briefly later  in this paper. The reader is advised to consult \cite{Nilsson:2013tva} for definitions and more details on the spin 3 sector of the higher spin system. The problematic issue of constructing source terms was mentioned in this context at the end of  that paper, and a suggestion  how it can be  solved is presented in the next section. 

The higher spin algebras together with the linearized versions of the zero field strength and unfolded equations have been discussed in many papers in the past, see, e.g., 
\cite{Shaynkman:2001ip, Blencowe:1988gj, Pope:1989vj, Horne:1988jf, Fradkin:1989xt, Vasiliev:1992ix, Vasiliev:2012vf}
and referencies therein. The conformal higher spin sector of the theory that is the subject of this paper is also analyzed in a recent paper by Vasiliev \cite{Vasiliev:2015mka}
where its relation to higher spin theory in $AdS_4$ is used to draw conclusions about Lagrangians etc. The scalar sectors, on the other hand, are not the same. In fact,
the scalar considered in this paper is  the one discussed in \cite{Shaynkman:2001ip}. The linearized spin 3 frame field system used in section 4 below  is 
discussed in the "metric" formulation in, e.g., \cite{Bergshoeff:2009tb}. Furthermore, the explicit analysis of the conformal higher spin system performed in this paper is closely related to the more formal
approach of $\si_-$ cohomology developed by Shaynkman and Vasiliev, see, e.g., \cite{Shaynkman:2000ts, Shaynkman:2004vu, Shaynkman:2014fqa}.

%%%%%%%%%%%%%%%%%%%%%%%%%%%%%%%%%%%%%%%%%%%%%%%%%%%%%%%%%%%%

\section{The conformal interacting higher spin equations}

%%%%%%%%%%%%%%%%%%%%%%%%%%%%%%%%%%%%%%%%%%%%%%%%%%%%%%%%%%%%

The two basic field equations for the $SO(3,2)$ conformal higher spin (HS) theory coupled to a scalar field with  fifth order self-interactions 
that we propose and study here are the {\it unfolded} equation
\beq
\label{unfoldeq}
{\mathcal D}\Phi|0\ra_q=S|0\ra_q,
\eeq
where ${\mathcal D}=d+A$, and the following {\it field strength} equation valued in the $SO(3,2)$  higher spin algebra 
\beq
\label{fteq}
F=T.
\eeq

%%%%%%%%%%%%%%%%%%%%%%%%%%%%%%%%%%%%%%%%%%%%%%%%%%%%%%%%%%%%

\subsection{The higher spin setup}

%%%%%%%%%%%%%%%%%%%%%%%%%%%%%%%%%%%%%%%%%%%%%%%%%%%%%%%%%%%%

We now explain the notation and content of these equations following \cite{Nilsson:2013tva}.
$F=dA+A\wedge A$ is the HS field strength obtained from the HS gauge field $A$ with the expansion
\beq
A=\Sigma_{n=1}^{\infty}(-i)^nA_n,\,\,\,\,A_n=e^{a_1....a_n}P_{a_1....a_n}+......+f^{a_1....a_n}K_{a_1....a_n}.
\eeq
To understand the structure  of this gauge field we give 
the parts of the higher spin conformal system that will explicitly play a role below, namely the spin 2 part
\beq
\label{spintwoA}
A_1=e^aP_a+\om^aM_a+bD+f^aK_a,
\eeq
where the generators of  translation, Lorentz, dilatation and special conformal transformations are, respectively, $P^a, M^a, D$ and $K^a$ with their associated 
gauge fields $e^a, \om^a, b$ and $f^a$, and the spin 3 part
\beq
\label{spinthreeA}
A_2=e^{ab}P_{ab}+\te^{ab}\tP_{ab}+\te^{a}\tP_{a}+\tom^{ab}\tM_{ab}+\tom^{b}\tM_{a}+\tb\tD+\tf^a\tK_a+\tf^{ab}\tK_{ab}+f^{ab}K_{ab}.
\eeq
The gauge fields (lower case quantities) and generators (upper case) of the HS algebra appearing in these expressions are all in irreps, i.e.,  the $a_1....a_n$
are totally symmetric and traceless sets of  three-dimensional vector indices. The fields $e^{a_1....a_n}$ are the spin $s=n+1$ frame fields and we will call
$f^{a_1....a_n}$ the  Schouten tensor\footnote{A perhaps more appropriate definition of the Schouten tensor is used, e.g.,  in \cite{Bergshoeff:2009tb}, which corresponds to the spin 3 field $\tf^{ab}$
in (\ref{spinthreeA}).}
 since $Df^{a_1....a_n}+....=0$ turns out to be the  spin $s=n+1$ Cotton equation
which is of order $2s-1$ in derivatives. We emphasize here that all the fields depend only on the three dimensional space-time coordinates $x^{\mu}$ and 
there are thus no dependence on any other coordinates or auxiliary variables like the ones\footnote{The notation adopted by Vasiliev often use
$y^{\al}$ which correspond to our $q^{\al}, p_{\al}$ while the auxiliary $z^{\al}$ "coordinates"  have no analogue here.} often appearing in  Vasiliev's constructions of 
interacting higher spin theories in $AdS$.

The HS algebra can be defined as follows. Consider the $so(2,1)\approx sp(2,{\bf R})$ spinor  variables $q^{\al}, p_{\al}$ (with $\al, \be,..=1,2$) which are 
hermitian operators satisfying  $[q^{\al}, p_{\be}]=i\de^{\al}_{\be}$. The spin $s=n+1$ HS generators are then given by
all Weyl ordered polynomials in $q^{\al}, p_{\al}$ of degree $2n$. For example, for $s=2$ we have
\beqa
&&P^a(2,0)=-\half (\si^a)_{\al\be}q^{\al}q^{\be},\,\,\,M^a(1,1)=-\half (\si^a)_{\al}{}^{\be}q^{\al}p_{\be},\\
&&D(1,1)=-\quart (q^{\al}p_{\al}+p_{\al}q^{\al}),\,\,\,K^a(0,2)=-\half (\si^a)^{\al\be}p_{\al}p_{\be}
\eeqa 
and for $s=3$ 
\beqa
P^{ab}(4,0)&=&\quart (\si^a)_{\al\be}(\si^b)_{\ga\de}q^{\al}q^{\be}q^{\ga}q^{\de},\\
\tP^{ab}(3,1)&=&\frac{1}{4}(\si^a)_{\al\be}(\si^b)_{\ga}{}^{\de}q^{\al}q^{\be}q^{\ga}p_{\de},\\
\tP^{a}(3,1)&=&\frac{1}{16}(\si^a)_{\al\be}(q^{\al}q^{\be}q^{\ga}p_{\ga}+q^{(\al}q^{\be}p_{\ga}q^{\ga)}+q^{(\al}p_{\ga}q^{\be}q^{\ga)}+p_{\ga} q^{\al}q^{\be}q^{\ga}),\,\,\,etc.
\eeqa
By computing the algebra of these generators keeping only single commutator terms  we obtain the classical higher spin algebra based on the Poisson bracket used in this context in the original work
on conformal higher spins in three dimensions \cite{Pope:1989vj}.
Instead, by quantizing the variables
$q^{\al}, p_{\al}$ and Weyl order them as above they generate the  for us relevant higher algebra  of  $SO(3,2)$. Note that for all generators $G(2n)$ (which are of order $2n$ in $q^{\al}, p_{\al}$) with $n$ vector indices
the ordering of the $q$ and $p$  operators do not matter and they are thus automatically ordered as required.

We give the operators  $q^{\al}$ dimension $L^{\half}$ and $p_{\al}$ dimension $L^{-\half}$ which means that both $A$ and 
$F$ will be dimensionless. This will be useful later when we discuss how to construct $S$ and $T$ on the RHSs of (\ref{unfoldeq}) and (\ref{fteq}).
With these rules all multiplications can be viewed as {\it star products} which, however,  has to be remembered since it is not explicitly shown by our notation.

We now turn to the LHS of (\ref{unfoldeq}). The derivative operator appearing there is just ${\mathcal D}=d+A$ where $A$ is as defined above\footnote{The derivative $D$ used  
frequently below is defined to contain only the spin 2 spin connection $\om^a$.}. However, the scalar field
$\Phi(x)$ is special and differs in its definition form $A$. $\Phi$ is expanded only in terms of the most special conformal generators $K^{a_1...a_n}$ which is the last one 
of the generators in each spin $s=n+1$ field $A_n$ above and contains {\it only} the variables $p_{\al}$, in fact, exactly $2n$ of them. We define the HS scalar field as follows
\beq
\Phi(x)=\Sigma_{n=0}^{\infty}(-i)^n\phi^{a_1...a_n}(x)K_{a_1...a_n},
\eeq
where the first term defines the usual scalar field $\phi(x)$ that will appear conformally coupled to spin 2 and all higher spin frame fields coming from $A$ and with its own fifth order self-interaction in the Klein-Gordon equation.

The vacuum used in (\ref{unfoldeq})  is 
defined to be annihilated by the $q^{\al}$ operators making it translationally invariant in the sense that $P^a|0\ra_q=0$. Although $\Phi$ itself does not contain any $q^{\al}$ operators, the fact that $A$ does  will lead to the appearance  of  
 interaction terms already for spin 2. In particular, a correctly normalized $R\phi$ interaction term appears directly after starting the unfolding procedure as observed in \cite{Nilsson:2013tva}. The scalar field is conformal 
and thus of dimension $L^{-\half}$ so the LHS of (\ref{unfoldeq}) is a one-form of dimension $L^{-\half}$ which must  be true also for $S$ on the RHS of that equation.
We will propose an expression for $S$ below after explaining the structure of the second equation $F=T$.

The role of the vacuum in the unfolded equation (\ref{unfoldeq}) is clear and a well-known property of this kind of scalar field, see, e.g.,  \cite{Shaynkman:2001ip}. 
However, one of the crucial points in this discussion is  to understand the relation of the two field equations (\ref{unfoldeq}) and (\ref{fteq}) where the former one involves the vacuum while  the latter one 
does not  and hence has components for every generator of the higher spin algebra. 
To make the following argument a bit more explicit we give the spin 2 and spin 3 equations 
coming from the generator decomposition of $F=0$. Note, however, that the following spin 2 and 3 equations have been truncated to the single commutator terms for simplicity.

 For spin 2 the equations are (in the gauge $b_{\mu}=0$)\cite{Horne:1988jf}
\beqa
\label{spintwoFzero}
F=0:\,\,\,&&T^a=0,\\
&&R^a-2\ep^a{}_{bc}e^b\wedge f^c=0,\\
&&e^a\wedge f_a=0,\\
&&Df^a=0,
\eeqa
where $T^a=De^a=de^a+\om^a{}_{bc}\om^b\wedge e^c$ and $R^a=d\om^a+\half \om^a{}_{bc}\om^b\wedge \om^c$. The second of these is useful for us since solving it for $f_{\mu}{}^a$ gives 
\beq
\label{fcotton}
f_{\mu\nu}=\half(R_{\mu\nu}-\frac{1}{4}g_{\mu\nu}R)=\half S_{\mu\nu},
\eeq
where $S_{\mu\nu}$ is the Schouten tensor. The third equation in the above $F=0$ spin 2 system then just says that the Schouten tensor is symmetric and the last equation that it satisfies the Cotton equation.

For spin 3 we give only the cascading equations (see the next section) used to express the spin 3 Schouten tensor $f_{\mu}{}^{ab}$ in terms of the frame field $e_{\mu}{}^{ab}$ (the full system including the constraint equations is discussed in \cite{Nilsson:2013tva, Linander:2015xx})
\beqa
\label{spinthreeFzero}
&& F^{ab}(4,0)=De^{ab}+e^c\wedge \te^{d(a}\ep_{cd}{}^{b)}-(e^{(a}\wedge \te^{b)}-trace)=0,\cr
&& F^{ab}(3,1)=D\te^{ab}-2e^c\wedge \tom^{d(a}\ep_{cd}{}^{b)}-(e^{(a}\wedge \tom^{b)}-trace)-4f^c\wedge e^{d(a}\ep_{cd}{}^{b)}=0,\cr
&& F^{ab}(2,2)=D\tom^{ab}+3e^c\wedge \tf^{d(a}\ep_{cd}{}^{b)}-(e^{(a}\wedge \tf^{b)}-f^{(a}\wedge \te^{b)}-trace)+3f^c\wedge \te^{d(a}\ep_{cd}{}^{b)}=0,\cr
&& F^{ab}(1,3)=D\tf^{ab}-4e^c\wedge f^{d(a}\ep_{cd}{}^{b)}+(f^{(a}\wedge \tom^{b)}-trace)-2f^c\wedge \tom^{d(a}\ep_{cd}{}^{b)}=0,\cr
&& F^{ab}(0,4)=Df^{ab}+(f^{(a}\wedge \tf^{b)}-trace)+f^c\wedge \tf^{d(a}\ep_{cd}{}^{b)}=0.
\eeqa
From the explicit structure of these equations (and the work in  \cite{Pope:1989vj, Nilsson:2013tva}) it should be clear that they can all  be solved algebraically except
for the very last one which is the Cotton equation and that this  works for all spins. The result of this procedure thus expresses the spin $s=n+1$ Schouten tensor
$f_{\mu}{}^{a_1...a_n}$ in terms of the frame field $e_{\mu}{}^{a_1...a_n}$, a relation that involves $2s-2$ derivatives. In order to be able to introduce interactions, i.e., a stress tensor on the RHS of all the Cotton equations for arbitrary spin we must relax the equation $F=0$ and instead consider $F=T$ where the RHS must have the property that
all  the field strengths $F^{a_1...a_n}(0,2n)$ pick up the proper source terms. 

Having  concluded that all the component equations $F(n_q,n_p)=0$ for $n_q>0$ can be solved we note that the field strength $F$ reduces to
\beq
F=\Sigma_{n=0}^{\infty}(-i)^nF^{a_1...a_n}(0,2n)K_{a_1...a_n},
\eeq
i.e., it has become a field with the same structure as $\Phi$ defined above. This reduction of $F$ may, however, not be compatible with the Bianchi identities.
Also
other parts of $F$ that are assumed zero here are probably only so  in the linear analysis performed in \cite{Nilsson:2013tva}. As will be elaborated upon elsewhere 
\cite{Linander:2015xx} some of the constraint equations  of the spin 3 system along other generators than $K^{ab}$ 
will contain the spin 2 Cotton tensor and will thus be affected by the introduction of source terms. Hence this description of the structure of $F$
implies that $F$ can not be made to act directly on the vacuum like in  the unfolded equation (\ref{unfoldeq})  since then we would loose information. This suggests
that the proper equation for $F$ and $T$ involving the vacuum is instead the integrability equation for (\ref{unfoldeq}) namely\footnote{ Note that applying another derivative gives 
an identity after using the Bianchi identity for $F$.}
\beq
F\Phi|0\ra_q={\mathcal D}S|0\ra_q,
\eeq
with $F$ taking values in the whole HS algebra, which  implies that
\beq
\label{tphiequalsds}
T\Phi|0\ra_q={\mathcal D}S|0\ra_q,
\eeq
from which it should be possible to construct $T$.

We have now explained how to view the two equations (\ref{unfoldeq}) and (\ref{fteq}) and also defined the RHSs of these equations without 
giving any explicit expressions for them.
The main result of this paper  is that it is in fact   possible  to construct the RHSs
producing a fully interacting theory with  back reactions in both equations. We emphasize here that we have not yet provided a complete proof that the equations we propose
constitute a  
consistent system. In this context it might also be relevant to analyze the equation (where $\star$ is the Hodge dual)
\beq
{\mathcal D}\star{\mathcal D}\Phi|0>={\mathcal D}\star S|0>,
\eeq
which follows directly from the unfolded equation and contains the Klein-Gordon equation already at level $n=0$. It seems that for these different equations derived from the unfolded equation to be consistent with one another,
the information of the original equation is merely reshuffled to different levels but otherwise the same. Under what conditions this is true (if at all) remains to be  demonstrated, however.

Turning to the source terms we start by constructing $S$. It was explicitly shown in \cite{Nilsson:2013tva} that the unfolded equation with a zero RHS gives rise to
the conformal interaction term $R\phi$ with the correct coefficient for three dimensions. The goal now is to construct a RHS such that also the fifth order interaction term
is generated after unfolding the equation. In fact, also the scalar terms in the full Cotton equation (\ref{spintwocotton}) require the addition of a source term. The only structures that can be written down  which are one-forms with dimension $L^{-\half}$ and could generate the wanted terms are in fact,
\beq
S=-i\lam_1M(\Phi^*\Phi)\Phi -i \lam_2 K(\Phi^*\Phi)^2 \Phi,
\eeq
where $\lam_1$ and $\lam_2$ are two  free parameters. Here  we have used the definitions
\beq
M=dx^{\mu}e_{\mu}{}^aM_{a},\,\,\,K=dx^{\mu}e_{\mu}{}^aK_{a},
\eeq
where $M_a$ and  $K_{a}$ are the spin 2 Lorentz and special conformal generators of dimension $L^0$ and $L^1$, respectively. Unfolding (\ref{unfoldeq}) indeed gives the correct Klein-Gordon equation
at the spin 2 level (see below) and interestingly enough also the correct Cotton equation. As described for spin 3 in \cite{Nilsson:2013tva} this unfolding can be carried out further up in spin without any problems. There are, however, features involving infinite sets of higher spin terms in the full equations which probably means that the equations have to be iterated and truncated at some desired high spin level.

However, for this to work in the sense of producing  the spin 2 Cotton equation with the correctly coupled  scalar field as in (\ref{spintwocotton}) one further step is required. As we will clear  below 
we have to make use of the possibility to shift the gauge fields in $A$ by tensor terms which for spin 2 we choose as
\beq
\hat A_1=A_1+\lam_1 M\phi^2.
\eeq
 We will, however, not work with this shifted gauge field but instead move the tensor term over to the RHS of the unfolded equation. Combining this  term with the one already in $S$ we find that the RHS becomes
 \beqa
 \label{ssources}
 S_M&=&-i\lam_1M(\phi^a\phi^b-2\phi\phi^{ab})K_{ab}+...)\Phi,\\
 S_K&=&-i\lam_2 K(\phi^4+2(\phi^a\phi^b-2\phi\phi^{ab})\phi^2K_{ab}+.....)\Phi.
 \eeqa
 Note that a corresponding term for the $P_a$ generator does not exist since the term $P$ in $A$ is of dimension zero so it cannot  contain any   factors of $\Phi$. 
 It may be mentioned in this context that the unfolded equation will itself produce the full spin 2 Cotton equation with the stress tensor as a source. For higher spins one may 
 speculate about the structure of the corresponding Cotton equations. For spin 3 for instance, the Cotton tensor is in the irrep ${\bf 7}$ and has dimension $L^{-4}$, compared to $L^{-3}$ for spin 2, and hence
 the two-scalar terms must contain one further derivative which should result from the unfolding. Also other more complicated terms are possible with  derivatives distributed
 between scalars and higher spin frame fields in various ways.  It is even possible that there is  a non-derivative $e_{\mu ab}\phi^{10}$ term as a source in the spin 3 Cotton equation which may come 
from further terms in the source $S$. E.g., one may envisage  terms containing $e_{\mu}{}^{ab}$ which must involve $e_{\mu}{}^{ab}M_{ab}$, $e_{\mu}{}^{ab}K_{ab}$,  etc, multiplied by  $|\Phi|^4\Phi$, $|\Phi|^{8}\Phi$, etc  for dimensional reasons. 
How it is possible for  such  terms in $S$  to affect the spin three equations  will be clear below. Note that the issue  of whether or not terms like these will contribute also to the Klein-Gordon equation   depends on
traces like $e^{\mu}{}_ae_{\mu}{}^{ab}$ being non-zero. However, this is probably not the case since they can be set to zero by  higher spin  "scale" transformations. 

In a similar manner we may deduce the structure of $T$ in the HS equation (\ref{fteq}). $T$ must be a two-form of zero dimension giving rise to, after unfolding,
both $\pd_{\mu}\phi\pd_{\nu}\phi$ and $\phi D_{\mu}\pd_{\nu}\phi$ type terms. An especially intriguing fact is that derivative terms of the  kind
$\phi D_{\mu}\pd_{\nu}\phi$  may only arise through unfolding. The $T$ that has these properties will not be presented here and we hope to come back to this question
elsewhere. Note that a structure similar to $S$, i.e., 
\beq
\label{trialt}
T=-i g_1 \star P(\Phi^{*}\Phi)-i g_2\star M (\Phi^{*}\Phi)^2-i g_3 \star K (\Phi^{*}\Phi)^3,
\eeq
where the two-forms are
\beq
\star P= \half dx^{\mu}\wedge dx^{\nu}\ep_{\mu\nu}{}^{\rho}e_{\rho}{}^aP_{a},\,\,\,etc.
\eeq
will not suffice since terms with explicit derivatives seems to be needed. In fact, this follows directly from (\ref{tphiequalsds}) which of course will imply relations between parameters in $T$ and $S$.
Nevertheless, it is the first term in (\ref{trialt})  that has the correct structure to generate the required source term for the spin 2 Cotton equation in $F=T$. As for $S$ also $T$ will contain 
HS contributions of the kind $e_{\mu}{}^{ab}P_{ab}$ etc.

%%%%%%%%%%%%%%%%%%%%%%%%%%%%%%%%%%%%%%%%%%%%%%%%%%%%%%%%%%%%

\subsection{Explicit unfolding}

%%%%%%%%%%%%%%%%%%%%%%%%%%%%%%%%%%%%%%%%%%%%%%%%%%%%%%%%%%%%

In order to perform some  checks we need to unfold the scalar equation 
\beq
\label{unfoldedeq}
{\mathcal D}\Phi|0\ra_q=S|0\ra_q,
\eeq
 to find expressions for some of the first terms in the expansion. The point we want to emphasize here is that this equation contains, apart from the scalar field equation,
also  the higher spin field equations obtained by solving the equation $F=0$ but now coupled to the scalar field.   We now demonstrate this explicitly by deriving  both the Klein-Gordon
equation  (\ref{kleingordon})  and the spin 2 Cotton equation  (\ref{cottoneq})
from the unfolded equation (\ref{unfoldedeq}).
 
We start by computing the first few levels of the left hand side of the unfolded equation.
 At level $n$ the expressions multiplying $K_{a_1...a_n}|0\ra_q$ are (where $D=d+\om(1,1)$ and ${\mathcal O}(HS)$ indicates further higher spin terms that can be computed when needed)
\beqa 
{\mathcal D}\Phi|0\ra_q:\,\,&&n=0:\,\,\,(\pd_{\mu}\phi+\phi_{\mu}+{\mathcal O}(e_{\mu}{}^{ab}\phi_{ab}(s=3)+...))|0\ra_q,\\
&&n=1:\,\,\,(D_{\mu}\phi^{a}+f_{\mu}{}^a\phi+6\phi_{\mu}{}^a+{\mathcal O}(s\geq 3))K_a|0\ra_q,\\
&&n=2:\,\,\,(D_{\mu}\phi^{ab}+f_{\mu}{}^{(a}\phi^{b)}+15\phi_{\mu}{}^{ab}+{\mathcal O}(s\geq 3))K_{ab}|0\ra_q,
\eeqa
where we need to keep in mind that the uncontracted flat indices are always in irreps, i.e., in symmetrized traceless representations.
This means that for levels $n\geq 1$ each equation splits into three irreducible parts $n-1$, $n$ and $n+1$ obtained by multiplying it  with the level one generators 
$P^{\mu}:=e^{\mu}{}_{ a}P^a$, $M^{\mu}:=e^{\mu}{}_{ a}M^a$ and $K^{\mu}:=e^{\mu}{}_{ a}K^a$,
respectively. We refer to  the resulting equations as $n^-,n^0,n^+$, respectively. Applying this procedure to the $n=1$ equation above we find
\beqa
n=1^-:\,\,\,&&D_{\mu}\phi^{\mu}+f_{\mu}{}^{\mu}\phi+{\mathcal O}(s\geq 3),\\
n=1^0:\,\,\,&&\ep^{\mu\nu a}(D_{\mu}\phi_{\nu}+f_{\mu\nu})+{\mathcal O}(s\geq 3),\\
n=1^+:\,\,\,&&D_{(\mu}\phi_{\nu)}+f_{(\mu\nu)}+6\phi_{\mu\nu}+{\mathcal O}(s\geq 3).
\eeqa
Setting ${\mathcal D}\Phi|0\ra_q=0$ we can insert the level $n=0$ result into these equations and find
\beqa
n=1^-:\,\,\,&&-\Box\phi+f_{\mu}{}^{\mu}\phi+{\mathcal O}(s\geq 3)=0,\\
n=1^0:\,\,\,&&\ep^{\mu\nu a}f_{\mu\nu}+{\mathcal O}(s\geq 3)=0,\\
n=1^+:\,\,\,&&-D_{(\mu}\pd_{\nu)}\phi+f_{(\mu\nu)}+6\phi_{\mu\nu}+{\mathcal O}(s\geq 3)=0.
\eeqa

At level 2 we find the following LHSs of the unfolded equation
\beqa
n=2^-:\,\,\,&&D_{\mu}\phi^{\mu a}+\half f_{\mu}{}^{\mu}\phi^a+\frac{1}{6}f^{a b}\phi_b+{\mathcal O}(s\geq 3),\\
n=2^0:\,\,\,&&\ep^{\mu\nu (a}D_{\mu}\phi_{\nu}{}^{b)}+\half \ep^{\mu\nu (a}f_{\mu}{}^{b)}\phi_{\nu}+{\mathcal O}(s\geq 3),\\
n=2^+:\,\,\,&&D_{(\mu}\phi_{\nu\rho)}+f_{(\mu\nu}\phi_{\rho)}+15\phi_{\mu\nu\rho}+{\mathcal O}(s\geq 3).
\eeqa

To get a feeling for the non-trivial information in these equations we again assume ${\mathcal D}\Phi|0\ra_q=0$ and continue by  analyzing the first of the $n=2$ equations, the $2^-$ one. To do that we need to
use information from the two lower levels. This gives (dropping $s\geq 3$ terms)
\beq
\frac{1}{6}D_{\nu}(D^{(\nu}\pd^{\mu)}\phi-\frac{1}{3}g^{\nu\mu}\Box\phi-f^{(\nu\mu)}\phi+
\frac{1}{3}g^{\nu\mu}f_{\rho}{}^{\rho}\phi)-\frac{1}{2}f_{\nu}{}^{\nu}D^{\mu}\phi-\frac{1}{6}f^{\mu\nu}\pd_{\nu}\phi=0,
\eeq
which simplifies to
\beq
\Box\pd_{\mu}\phi-\frac{1}{3}D_{\mu}\Box\phi-(D_{\nu}f_{\mu}{}^{\nu})\phi-2f_{\mu}{}^{\nu}\pd_{\nu}\phi+
\frac{1}{3}(D_{\mu}f_{\nu}{}^{\nu})\phi-\frac{8}{3}f_{\nu}{}^{\nu}\pd_{\mu}\phi=0.
\eeq
Then using the fact (where the  zero torsion condition is assumed)
\beq
\Box\pd_{\mu}\phi=D_{\mu}\Box\phi+R_{\mu}{}^{\nu}\pd_{\nu}\phi
\eeq
and the Klein-Gordon equation, we find the above $2^-$ equation  to read
\beq
(R_{\mu\nu}-2f_{\mu\nu}-2f_{\rho}{}^{\rho}g_{\mu\nu})\pd^{\nu}\phi-(D_{\nu}f_{\mu}{}^{\nu}-D_{\mu}f_{\nu}{}^{\nu})\phi=0.
\eeq
We note then that this equation becomes an identity if we set
\beq
\label{fschouten}
f_{\mu\nu}=\half S_{\mu\nu}=\half (R_{\mu\nu}-\frac{1}{4}g_{\mu\nu}R),
\eeq
where $S_{\mu\nu}$ is the Schouten tensor. As we have seen above in (\ref{fcotton}), the equation $F=0$ also contains this information.

The equation $n=1^+$ above can now be seen to play an interesting role in rewriting the Cotton equation in a way that will help us  to guess source terms for the entire higher spin system.
The point is that if we use (\ref{fschouten}) in the $n=1^+$ equation we can  eliminate the term $D_{(\mu}\phi_{\nu)}=-D_{\mu}\pd_{\nu}\phi$ from the Cotton equation in (\ref{spintwocotton}). This gives 
\beq
\label{topgaugedcotton}
C_{\mu\nu}=\frac{3g}{8}(\phi_{\mu}\phi_{\nu}-2\phi\phi_{\mu\nu})|_{\bf 5},
\eeq
which is the form of the Cotton equation we will now show can be derived from the unfolded equation in (\ref{unfoldedeq}).

Now we turn to the second of the $n=2$ level equations, the $2^0$ in the irrep ${\bf 5}$. Making use of the lower level equations it becomes
\beq
\ep^{\mu\nu(a}(D_{\mu}D_{\nu}\pd^{b)}\phi-(D_{\mu}f_{\nu}{}^{b)})\phi-2f_{\mu}{}^{b)}\pd_{\nu}\phi)=0.
\eeq
Using the Ricci identity this equation reads
\beq
\ep^{\mu\nu(a}(R_{\mu}{}^{b)}\pd_{\nu}\phi-(D_{\mu}f_{\nu}{}^{b)})\phi-2f_{\mu}{}^{b)}\pd_{\nu}\phi)=0,
\eeq
and setting  $f_{\mu\nu}=\half S_{\mu\nu}$ as found above it reduces to 
\beq
-\half C^{ab}\phi=0,
\eeq
where $C^{ab}=\ep^{\mu\nu(a}D_{\mu}R_{\nu}{}^{b)}$ is the Cotton tensor. Again we find information present also in the equation $F=0$. 
Thus is it clear that while the $F=0$ equations contain, of course, only higher spin dynamics without scalar field sources the unfolded equation ${\mathcal D}\Phi|0\ra_q=0$
contains dynamical information for both the scalar field and the higher spin fields but without any non-trivial couplings between the scalar field  and the higher spin ones.
Introducing sources must thus be done for  both equations in a consistent way. We will address  this issue again below. 

We now introduce the  non-zero source terms  (\ref{ssources}):
\beq
{\mathcal D}\Phi|0\ra_q=(S_M+S_K)|0\ra_q.
\eeq
One crucial property of this expression for the source is that it is zero at level $n=0$, and that $S_M$ contributes only to the $n^0$  equations at level $n$  while $S_K$ contributes only to the $n^-$ equations. Thus there are no source terms affecting the $n^+$ equations which therefore are the same as for ${\mathcal D}\Phi|0\ra_q=0$ where it is used to  determine the fields $\phi^{a_1...a_{n+1}}$ in terms of fields at lower levels. This is seen as follows:
consider a general term at level $n$ in the expansion of $K_{\mu} \Phi^5|0\ra_q=e_{\mu}{}^aK_{a} \Phi^5|0\ra_q$ which we write as 
$e_{\mu}{}^a(\Phi^5)^{b_1...b_n}K_{ab_1...b_{n-1}}|0\ra_q$.
Contracting it with $P^{\mu}$ then gives  $K_{b_1...b_{n-1}}$. Contraction with $M^{\mu}$ gives instead $e_{\mu}{}^a\ep^{\mu}{}_{(a}{}^cK_{b_2...b_{n-1})c}=0$ and using $K^{\mu}$
one gets $e^{\mu}{}^aK_{\mu ab_1...b_{n-1}}=0$. $S_M$ works in a similar way with contributions only to the $n^0$ equations.

We will now continue to analyze the effects of adding the explicit source terms given in (\ref{ssources}):
 \beqa
 S_M&=&-i\lam_1M(\phi^a\phi^b-2\phi\phi^{ab})K_{ab}+...)\Phi,\\
 S_K&=&-i\lam_2 K(\phi^4+2(\phi^a\phi^b-2\phi\phi^{ab})\phi^2K_{ab}+.....)\Phi.
 \eeqa
Using the results
\beqa
&&[M_a,K_{bc}]=-2i\ep_{a(b}{}^dK_{c)d},\\
&&M^a[M_a,K_{bc}]|0\ra_q=-12K_{bc}|0\ra_q,
\eeqa
the RHS of
the unfolded equation is at the first few levels
\beqa
M^{\mu}(S_M)_{\mu}|0\ra_q:&&\,\,\,\,=0,\,\,\,\,n=0,\\
&&\,\,\,\,=0,\,\,\,\,n=1^0,\\
&&\,\,\,\,=-12\lam_1 (\phi^a\phi^b-2\phi\phi^{ab})\phi K_{ab}|0\ra_q,\,\,\,\,n=2^0.
\eeqa
In the case of $S_K$  we need the results
\beq
[P_a,K_b]=-2i\ep_{ab}{}^cM_c-2i\eta_{ab}D \Rightarrow [P^a,K_a]|0\ra_q=-3|0\ra_q,
\eeq
\beq
[P_a,K_{bc}]|0\ra_q=-6(\eta_{a(b}K_{c)}-\tfrac{1}{3}\eta_{bc}K_a)|0\ra_q \Rightarrow [P^a,K_{ab}]|0\ra_q=-10K_b|0\ra_q.
\eeq
We find
the following  contributions  to the $n^-$ equations 
\beqa
P^a6\lam K_a \Phi^5|0\ra_q&&=0,\,\,\,\,n=0,\\
&&=-3\lam_2\phi^5|0\ra_q,\,\,\,\,n=1^-,\\
&&=-10\lam_2 \phi^b \phi^4 K_{b}|0\ra_q,\,\,\,\,n=2^-.\\
%&&=P^a30\lam (\phi^{bc}\phi^4+4\phi^b\phi^c \phi^3) K_{abc}|0\ra_q,\,\,\,\,n=3,
\eeqa

With these results for the source terms  of the unfolded equation 
\beq
{\mathcal D}\Phi|0\ra_q=(S_M+S_K)|0\ra_q,
\eeq 
the first few level equations become, dropping terms involving higher spin fields,
\beqa
n=0:&&\,\,\,\,\phi_{\mu}=-\pd_{\mu}\phi,\\
n=1^-:&&\,\,\,\,\Box\phi-\tfrac{1}{8}R\phi-3\lam_2\phi^5=0,\\
n=1^0:&&\,\,\,\,0=0,\\
n=1^+:&&\,\,\,\,6\phi_{\mu\nu}=(D_{(\mu}D_{\nu)}\phi-f_{\mu\nu}\phi)|_{\bf 5},\\
n=2^-:&&\,\,\,\,(R_{\mu\nu}-2f_{\mu\nu}-2f_{\rho}{}^{\rho}g_{\mu\nu})\pd^{\nu}\phi-(D_{\nu}f_{\mu}{}^{\nu}-D_{\mu}f_{\nu}{}^{\nu})\phi=0,\\
n=2^0:&&\,\,\,\,C_{\mu\nu}=36\lam_1 (\phi_{\mu}\phi_{\nu}-2\phi\phi_{\mu\nu})|_{\bf 5},\\
n=2^+:&&\,\,\,\,\phi_{\mu\nu\rho}=-\tfrac{1}{15}(D_{(\mu}\phi_{\nu\rho)}+f_{(\mu\nu}\phi_{\rho)})|_{\bf 7}.
\eeqa
where we have adopted the solution $f_{\mu\nu}=\half S_{\mu\nu}=\half (R_{\mu\nu}-\frac{1}{4}g_{\mu\nu}R)$ to the $n=2^-$ equation in order to obtain the Cotton equation from 
the $n=2^0$ equation.
The Cotton equation obtained this way has the same structure as the one we were
seeking namely the one given in (\ref{topgaugedcotton})  so this seems to be on the right track. A crucial further test is to construct  the source $T$ in the adjoint equation $F=T$ such that
the same Cotton equation results as discussed above after equation (\ref{trialt}).  This should be possible to do and we hope to come back to this in a future publication.

We should also emphasize another feature of the calculation leading to these results. The fact that the scalar self-interaction term $K|\Phi|^4\Phi$
gives rise to new terms in both the $1^-$ and $2^-$ equations leads to a consistency check in the sense that
hese terms are seen to cancel in the the $2^-$ equation and hence  the result quoted for $f_{\mu\nu}$ 
is not affected by the addition of  the $K|\Phi|^4\Phi$ term.

%%%%%%%%%%%%%%%%%%%%%%%%%%%%%%%%%%%%%%%%%%%%%%%%%%%%%%%%%%%%

\section{A cascading Lagrangian}

%%%%%%%%%%%%%%%%%%%%%%%%%%%%%%%%%%%%%%%%%%%%%%%%%%%%%%%%%%%%

In this section  we will make use of a feature of the Chern-Simons gauge theory for the higher spin gauge field $A$ that allows us to show that
the component Lagrangian is naturally  expressed  in terms of the generalized spin connections as suggested in  \cite{Nilsson:2013tva}.
We will demonstrate explicitly that it is possible to derive such a Lagrangian once a certain subset of the $F=0$ component equations are solved. 
This subset of equations, which will be called cascading, does not include the spin $s$ Cotton equations which are therefore obtained by varying the resulting 
Lagrangian with respect to the frame fields for each spin $s\geq 2$. $F=0$ contains a number of other equations that one must use to determine other fields
contained in $A$ or prove are identities; for a complete discussion of the spin three situation see \cite{Nilsson:2013tva}.

We use here the  results coming from  the single commutator terms in the spin 2 and 3 cases as  examples of the technique. However,  these examples 
make it plausible   that
this procedure  works for all spins and in the full star product formulation where all multi-commutators and non-linearities are included.
%\footnote{The infinite number of terms in these equations will require the use of star product methods as is standard in $AdS$ higher spin theories.}
 Its origin is in the gauge Chern-Simons theory
\beq
S=\half Tr\int(AdA+\frac{2}{3}A^3),
\eeq 
where the trace is in the higher spin algebra which should generate   precisely the terms in the Lagrangian used in the cascading procedure described below. This Lagrangian can in principle 
be written out explicitly in terms of all the fields appearing in $A$ \cite{Fradkin:1989xt}. This will produce very complicated expressions and seems useful only for lower spin truncations. To get from this first order formulation
to a "second" order one in terms of only the frame fields seems even more complicated and any kind of simplifications that can be utilized in this context would be welcome. 
Below we will describe one potentially useful feature of this kind.
  
The standard spin 2 Chern-Simons like action reads in terms of $\om_1:=\om(1,1)$
\beq
S_2=\thalf \int (\om_1 d \om_1 + \tfrac{2}{3}\om_1 \wedge \om_1 \wedge \om_1),
\eeq
which leads to the field equation $C_{\mu\nu}=0$ for the Cotton tensor $C_{\mu\nu}=\ep_{\mu}{}^{\al\be}D_{\al}(R_{\be\nu}-\frac{1}{4}g_{\be\nu}R)$.
Here we use the notation from \cite{Nilsson:2013tva} for the $s=2$ spin connection obtained by solving the zero torsion condition.
In that paper it was suggested that this spin 2 action has a generalization to arbitrary spin in the sense that the action is naturally expressed in terms of
$\tom_n:=\tom(n,n)$ which is  a one-form in the irrep $n$ of $SO(1,2)$. (A spin 3 example is $\tom^{ab}$ appearing in the expansion of $A_2$ in (\ref{spinthreeA}).) 
%The final all-spin action then contains interaction terms between these generalized spin connections
%due to  multi-commutators or equivalently the star product formulation of the theory as explained briefly in \cite{Nilsson:2013tva}. 
%However, it is clear that
%the interaction terms will  contain other fields than the spin connections in the expansions of $A_n$. 

Here we will show how to derive the action for spin 3 which is of the suggested form 
\beq
\label{omegathreeaction}
S_3= \thalf \int \tilde \om_2 D \tilde \om_2 = \thalf \int (\tilde \om_2 d \tilde \om_2+ \tom_2 \wedge \om_1 \wedge \tom_2).
\eeq
Note that a cubic term with three $\tom_2$ does not exist. However, once the multi-commutators are taken into account there will appear new 
interaction terms  containing spin connections  with arbitrarily high spin.  Also other higher spin fields will occur in these interaction  terms and to  be clear about 
which fields we talk about  we consider again the spin 3 higher spin gauge field in (\ref{spinthreeA})
\beq
A_2=e^{ab}P_{ab}+\te^{ab}\tP_{ab}+\te^{a}\tP_{a}+\tom^{ab}\tM_{ab}+\tom^{b}\tM_{a}+\tb\,\tD+\tf^a\tK_a+\tf^{ab}\tK_{ab}+f^{ab}K_{ab}.
\eeq
Solving the first four of the equations in (\ref{spinthreeFzero}) will result in a cascading sequence of relations that will express
the one-form field $f^{ab}$, called the spin 3 Schouten tensor, in terms of the frame field $e^{ab}$ each step producing a new derivative.
The last equation in (\ref{spinthreeFzero}) is then the spin 3 Cotton equation containing five derivatives. The action we derive here uses only
the field $\tom^{ab}$ in $A_2$ denoted $\tom_2$ in (\ref{omegathreeaction}). In fact, as we will see below also the field $\tb$ in (\ref{spinthreeA}) will appear in the action 
but it turns out that this field  is expressible in terms 
of $\tom_2$ as shown in \cite{Nilsson:2013tva}.

The main goal of this section is to give a simple procedure for deriving a Lagrangian that gives the full non-linear Cotton equation
for any spin, and indeed for the whole higher spin system. This result follows provided some basic conditions to be specified below are met.
 Here we give the main ideas and  the details only for spin 2 and 3 but
it is likely  that this method can be generalized to the whole higher spin theory. 

%Decomposing this equation, we get the cascade equations which read
%\beqa
%&& De^{ab}+e^c\wedge \te^{d(a}\ep_{cd}{}^{b)}-(e^{(a}\wedge \te^{b)}-trace)=0,\\
%&& D\te^{ab}-2e^c\wedge \tom^{d(a}\ep_{cd}{}^{b)}-(e^{(a}\wedge \tom^{b)}-trace)-4f^c\wedge e^{d(a}\ep_{cd}{}^{b)}=0,\\
%&& D\tom^{ab}+3e^c\wedge \tf^{d(a}\ep_{cd}{}^{b)}-(e^{(a}\wedge \tf^{b)}+f^{(a}\wedge \te^{b)}-trace)+3f^c\wedge \te^{d(a}\ep_{cd}{}^{b)}=0,\\
%&& D\tf^{ab}-4e^c\wedge f^{d(a}\ep_{cd}{}^{b)}+(f^{(a}\wedge \tom^{b)}-trace)-2f^c\wedge \tom^{d(a}\ep_{cd}{}^{b)}=0,\\
%&& Df^{ab}+(f^{(a}\wedge \tf^{b)}-trace)+f^c\wedge \tf^{d(a}\ep_{cd}{}^{b)}=0.
%\eeqa
%
%These equations are easily solved and seen to give, in a step by step manner, the last field $f^{ab}$, which we call the spin 3 Schouten tensor, 
%in terms of the first one $e^{ab}$ called the spin 3 frame field. The last equation above is therefore the spin 3 Cotton equation and every field that occurs in the 
%expansion of the $so(3,2)$ gauge field $A_2$ can be expressed in terms of this frame field. In proving this last fact also the equations coming form $F=0$ but not given
%above must be solved or proven to be identities. This was demonstrated in \cite{Nilsson:} and the results from that analysis will be used here. The reader is advised to 
%consult that paper for further details.

We will now show how one can derive an action that automatically gives rise to the fifth order Cotton equation for this spin 3 system. 
In order to streamline the discussion we simplify the spin 3 equations in (\ref{spinthreeFzero}) as far as possible without destroying features of the system that are relevant
for this particular discussion.
First we note that the spin 3 equations contain  fields from the spin 2 system that we can discard at this point but put back if a  complete
analysis is required. This statement applies to all terms containing the spin 2 fields $\om^a$ and $f^a$  but not to the terms with a dreibein $e^a$.
This reduces the equations to
\beqa
&& de^{ab}+e^c\wedge \te^{d(a}\ep_{cd}{}^{b)}-(e^{(a}\wedge \te^{b)}-trace)=0,\\
&& d\te^{ab}-2e^c\wedge \tom^{d(a}\ep_{cd}{}^{b)}-(e^{(a}\wedge \tom^{b)}-trace)=0,\\
&& d\tom^{ab}+3e^c\wedge \tf^{d(a}\ep_{cd}{}^{b)}-(e^{(a}\wedge \tf^{b)}-trace)=0,\\
&& d\tf^{ab}-4e^c\wedge f^{d(a}\ep_{cd}{}^{b)}=0,\\
&& df^{ab}=0.
\eeqa

We now need to make use of the possibility to gauge fix the higher spin symmetries to further simplify these equations. 
As explained in \cite{Nilsson:2013tva} the field $\te^a$ can be set to zero by using the symmetries related to the parameters
$\tLam^{ab}(2,2)$, $\tLam^{a}(2,2)$, and $\tLam(2,2)$. This sets to zero the last term in the first equation above. 
 In order to eliminate also the last  term in the second and third equations we need to be able to choose a gauge
where $\tom^a=e^a\hat \om$ and $\tf^a=e^a\hat f$. However, while this is possible for $\tom^a$ it is not so for $\tf^a$. In the former case 
we can use $\tLam^{ab}(1,3)$ and $\tLam^{a}(1,3)$ to establish this fact but in the latter case we have only $\tLam^{ab}(0,4)$  at our disposal which means that 
the best we can do is to gauge fix to
\beq
\tf_{\mu}{}^a=\ep_{\mu}{}^{ab}\hat f_b+e_{\mu}{}^a \hat f,
\eeq
which unfortunately will complicate the situation somewhat.

Instead of trying to construct a Lagrangian directly for the frame fields and then perform a variation with respect to  the frame fields to obtain the Cotton equations these can be obtained in a manner that 
is slightly easier if we make use of the description of this system as a Chern-Simons gauge theory for the conformal group $SO(3,2)$.
In order to see how this is done we consider first the spin 2 Chern-Simons system which is  given in terms of the gauge field
\beq
A_1=e^aP_a+\om^aM_a+bD+f^aK_a,
\eeq
where the $SO(3,2)$ generators $P_a,M_a,D,K_a$ have been assigned one gauge field each.
The exercise is then to solve the zero field strength equation $F_1=0$ which if decomposed along the different generators become
 (here the Riemann tensor is $R^a=d\om^a+\half \ep^a{}_{bc}\om^b\wedge \om^c$ and we have imposed the gauge $b=0$)
\beqa
\label{spintwofzero}
&&T^a=de^a+\ep^a{}_{bc}\om^b\wedge e^c=0,\cr
&&R^a-2\ep^a{}_{bc}e^b\wedge f^c=0,\cr
&&Df^a=0,
\eeqa
which we call the cascading equations while the remaining equation
\beq
e^a\wedge f_a=0,
\eeq
is a constraint on the solution of the cascading system above. The first equation is solved for the spin connection in terms of the dreibein, the second for $f^a$ in terms of the Riemann tensor with the result that $f_{\mu}^ae_{\nu a}$ is just the symmetric Schouten tensor. The last equation is then a constraint that is automatically satisfied while the last 
of the cascading equations  becomes the Cotton equation. The goal is now to use these cascading equations to show that the variation of the action gives the Cotton equation. 

We start the cascading procedure from
\beq
L_1=-2e^a\wedge Df_a.
\eeq 
The variation of $L_1$ is 
\beq
\de L_1=-2\de e^a\wedge Df_a-2e^a \wedge D\de f_a-2e^a\wedge \ep_{abc}\de\om^b\wedge f^c,
\eeq
 which would give the Cotton equation by demanding $\de L_1=0$ if the last two  terms could be gotten rid off.
To achieve  this we note  first that the second term vanishes due to the torsion constraint after an integration by parts. To deal with the last term we 
add the standard Chern-Simons term
\beq
L_2=\thalf\om^a\wedge d \om_a + \tfrac{1}{6}\ep^{abc}\om_a \wedge \om_b \wedge \om_c,
\eeq
whose  variation is
\beq
\de L_2=\de \om_a \wedge R^a=2\de\om^a\wedge \ep_{abc}e^b\wedge f^c,
\eeq
where we have  used  the second equation in (\ref{spintwofzero}) in the last equality. Thus we obtain the
 Cotton equation as a result of varying the Lagrangian $L=L_1+L_2$. However, $L_1=0$ after an integration by parts as a consequence of the torsion constraint which is assumed solved in this analysis.
 This implies that the Lagrangian $L_2$ alone provides the Cotton equation when varied with respect to the dreibein $e_{\mu}{}^a$. This derivation of the Cotton equation is
 a bit too trivial to be interesting but for spin 3 and higher it seems to simplify the calculation of the spin s Cotton equation quite a bit. Recall that these equations 
 are of order $2s-1$ in derivatives.

We now turn to the spin 3 system and repeat these steps. To this end we note that the spin 3 Cotton equation in the simplified version given above follows trivially by varying the  Lagrangian
\beq
L_1= e^{ab}\wedge df_{ab},
\eeq
with respect to the explicit frame field. 
However,  this conclusion is only correct  if we can eliminate the second term in its  variation
\beq
\de L_1=\de e^{ab}\wedge d f_{ab}+e^{ab}\wedge d\de f_{ab}.
\eeq
But this can be done by  adding another  term to the Lagrangian whose variation cancels the last unwanted term.
The  term we need to add is 
\beq
L_2=e^c\wedge \te^{da}\ep_{cd}{}^b\wedge f_{ab}.
\eeq
The reason this works is that  in the variation
\beq
\de L_2= e^c\wedge \te^{da}\ep_{cd}{}^b \wedge \de f_{ab}+e^c\wedge\de \te^{da}\ep_{cd}{}^b\wedge f_{ab},
\eeq
the first term equals $-de^{ab}\wedge \de f_{ab}$ by using the field equation for $e^{ab}$ coming from $F=0$ (recall that we are in a gauge where $\te^a=0$). To make use of this field equation is of course allowed here since
it is algebraic and actually solved so that it is identically satisfied. This fact can now be used for all the "field equations" in $F=0$ except the last one which is the 
five derivative Cotton equation.

Having established this cancelation we now need to add a further term to cancel also the second term in $\de L_2$ above. The required term is
\beq
L_3=-\frac{1}{4}\te^{ab}\wedge d \tf_{ab},
\eeq
whose variation can be written, again making  use of the algebraic "field equations" this time the ones for $\tf^{ab}$ and $\te^{ab}$, as
\beq
\de L_3=-\frac{1}{4}\de \te^{ab}\wedge d \tf_{ab}-\frac{1}{4}\te^{ab}\wedge d \de \tf_{ab}=\de \te^{ad} \wedge e^c\wedge f_{ab}\ep_{cd}{}^{b}-\half e^c\wedge \tom^{da}\ep_{cd}{}^b\wedge \de \tf_{ab}.
\eeq
The remaining term is then the last one in the previous equation which we cancel by adding
\beq
L_4=\half e^c\wedge \tom^{da}\wedge \tf_{ab}\ep_{cd}{}^b,
\eeq
which varies into
\beq
\de L_4=-\half e^c\wedge \de\tom_{ab}\wedge \tf^{da}\ep_{cd}{}^b+\half e^c\wedge \tom^{da}\wedge \de \tf_{ab}\ep_{cd}{}^b.
\eeq
After canceling the last term against the same  term coming from $\de L_3$ we are left with the first term in $\de L_4$ which we write as
\beq
\label{lfourrest}
-\frac{1}{6}\de \tom^{ab}\wedge (d\tom_{ab}-e_a\wedge \tf_b).
\eeq
The first term in this expression is canceled by the variation of
\beq
L_5=\frac{1}{12}\tom^{ab}d\tom_{ab}.
\eeq
 Now we can use  the algebraic equations from $F=0$ again to find that $L_1+L_2=0$ and $L_3+L_4=0$. Since the procedure stops here  $L_5$
is actually  (the main part of) the final answer and  is precisely  the Lagrangian proposed in \cite{Nilsson:2013tva}. 

We have, however, still one term that we need to deal with, namely the second term in (\ref{lfourrest}), which as we will now see is of a slightly different nature.
We start by adding
\beq
L_6=-\frac{1}{6}\tom^{ab}\wedge e_a\wedge \tf_b.
\eeq
Varying  $\tom$  gives the term we need to cancel  in (\ref{lfourrest}) leaving us with the term
\beq
\label{tomedetf}
-\frac{1}{6}\tom^{ab}\wedge e_a\wedge \de \tf_b.
\eeq
Now we must consult \cite{Nilsson:2013tva} where it was shown that the one-forms $\tf^a$ and $\tb$ are given by
\beq
\tf_{\mu}{}^a=\tf_{[\mu\nu]}e^{\nu a}+e_{\mu}{}^a \hat f=-\frac{3}{8}\pd_{[\mu}\tb_{\nu]}+e_{\mu}{}^a \hat f, \,\,\,\,\tb_{\mu}=\frac{1Ö}{4}\tom_{\nu \mu}{}^{\nu}.
\eeq
In the present analysis where we keep track of only the spin 3 fields, the relevant expression we need is
$
\de\tf_{[\mu\nu]}=-\frac{3}{8}\pd_{[\mu}\de\tb_{\nu]}
$
which implies that we can rewrite   (\ref{tomedetf}) as
\beq
-\frac{1}{6}\tom^{ab}\wedge e_a\wedge \de \tf_b= \frac{1}{8}\tb\wedge  d\de\tb.
\eeq
Thus  the last spin 3 term to add is
\beq
L_7=-\frac{1}{16}\tb\wedge  d\tb.
\eeq

We have therefore shown that the spin 3 part of the  Lagrangian reads
\beq
L=L_5+L_6+L_7=\frac{1}{12}\tom^{ab}\wedge d\tom_{ab}-\frac{1}{6}\tom^{ab}\wedge e_a\wedge \tf_b- \frac{1}{16}\tb \wedge d\tb.
\eeq
The second term on the RHS is  related (see above) to the last one and we find the final form of the Lagrangian to be
\beq
L=\frac{1}{12}\tom^{ab}\wedge d\tom_{ab}+ \frac{1}{16}\tb\wedge  d\tb,
\eeq
which is therefore expressed entirely in terms of the spin connection $\tom^{ab}$ of the spin 3 sector.

One also needs to verify that the remaining constraint equations are satisfied as explained in \cite{Nilsson:2013tva}. 
It would then be interesting to see how the  different steps  in the cascading procedure are affected by increasing the spin, adding non-linear terms and coupling the system to other fields.
For the spin  2 - spin 3 system these questions may be  answered by the analysis of the full non-linear equations in \cite{Linander:2015xx}.
As noted  previously in this section the cascading trick  is suggested by writing out the original 
Chern-Simons action for $A$ using the trace over 
the higher spin algebra at each spin level separately as done in \cite{Fradkin:1989xt}. 
%%%%%%%%%%%%%%%%%%%%%%%%%%%%%%%%%%%%%%%%%%%

\section{Conclusions}

%%%%%%%%%%%%%%%%%%%%%%%%%%%%%%%%%%%%%%%%%

In this paper we have continued the approach to conformal higher spin theories in three dimensions set up recently in \cite{Nilsson:2013tva}.
There it was emphasized that the unfolded equation ${\mathcal D}\Phi|0\ra_q=0$ for the higher spin algebra based on $SO(3,2)$, 
the conformal group in three dimensional space-time, realized in terms of
two hermitian spinor operators $q^{\al},p_{\al}$ satisfying $[q^{\al},p_{\be}]=i\de^{\al}_{\be}$, produces the correct Klein-Gordon equation
for a conformal scalar coupled to the spin 2 metric and its generalization to spin 3\footnote{Note that there are other spin 3 terms in the Klein-Gordon equation than the one presented in \cite{Nilsson:2013tva}.}. 

Here we take this approach some steps further by proposing an unfolded equation for a scalar field
coupled to all higher spins $\geq 2$ including the $\phi^5$ self-interaction term in the Klein-Gordon equation. The expected scalar interactions with the  spin 2 and higher spin fields show also up in the field  equations and  are produced in the unfolding.
That the correct  spin 2 Cotton equation is obtained is checked explicitly while for spin 3 the corresponding equation can easily be  derived from this setup but it needs to be  checked independently.  Such a check would strengthen 
the argumentation for the higher spin equations suggested here.

We also present a simple method by which the Lagrangian for each higher spin field can be derived starting from the  $F=0$ equation valued in the higher spin algebra. This is demonstrated in section 3 for a truncated version of the spin 2 and 3 equations
and can be seen to be a consequence of expanding the Chern-Simons gauge theory action  using  the trace over the entire higher spin algebra. Then a cascading 
trick leaves the whole action written in a form where the  spin connections for each spin play a central role. The spin connections are here expressed as  $s-1$ derivatives acting on the spin $s$ frame field implying that ${\mathcal L_n}=\om_n d \om_n$ for $s=n+1$ contains $2s-1$ derivatives as it should. The result is a 
 "second" order formalism type Chern-Simons Lagrangian generalizing the standard one ${\mathcal L_1}(\om_1(e))$ for spin 2  to all spins. The cascading is here only performed for the linearized theory but will most likely   give the full answer once all interaction terms are included. 
 
 This approach also suggests a way to write down a higher spin Lagrangian in the higher spin language. One may try to combine the gauge Chern-Simons action 
 $S=\half Tr\int( AdA+\frac{2}{3}A^3)$ discussed in the previous paragraph (and in section 3) with other expressions 
  like
 (the star $\star$ is here a Hodge dual while the star product is implicit)
  \beq
 S=\int {}_p\la 0|\tilde\Phi^*(q){\mathcal D}\star {\mathcal D}\Phi(p)|0\ra_q,
 \eeq
 where we have introduced a  dual scalar
  field $\tilde\Phi(q)$ which is expanded in terms of even powers of $q^{\al}$ instead of $p_{\al}$ as for the ordinary scalar field $\Phi(p)$.
 By assumption the dual field $\tilde\Phi(q)=\phi(x)+\tilde\phi_{a}(x)P^a+...$ where $P^a=-\half (\si^a)_{\al\be}q^{\al}q^{\be}$  is non-zero on the vacuum ${}_p\la 0|$ which is used to produce a well-defined  inner product ${}_p\la 0|0\ra_q=1$.

This action can be expanded in component fields whose field equations   should correspond to  the scalar field equation ${\mathcal D}\star {\mathcal D}\Phi(p)|0\ra_q=0$. The source $S$ can probably be
hidden in the covariant derivative.
The unfolded equation ${\mathcal D}\Phi(p)|0\ra_q=0$ could then be regarded as a solution to this equation
 which  means that some information is lost if one instead solves only ${\mathcal D}\star {\mathcal D}\Phi(p)|0\ra_q=0$. It would be nice to have an action principle that directly generates the unfolded equation
 as the field equation.
 Interaction terms for the scalar field may also arise by considering actions of the kind
 \beq
 S_6=\int {}_p\la 0|(\tilde\Phi^*(q))^m\star (\Phi(p))^n|0\ra_q|_{m+n=6}.
 \eeq
  The actions considered here are dimensionless and the integrands are  three-forms but they seem not to produce in a simple way the  field equations used  previously
 in this paper. The main reason for this is that although the dual field $\tilde\Phi(q)$ is here assumed to start with $\phi(x)$ the terms of higher order in $q^{\al}$  will probably be very complicated 
 (even non-local). 
 
 Another potentially interesting aspect arises if this higher spin theory can be generalized to contain the topologically gauged spin theories derived in 
 \cite{Gran:2008qx, Chu:2009gi, Gran:2012mg}. Then
 perhaps the background solutions found in \cite{Chu:2009gi, Nilsson:2013fya} could be lifted to the higher spin theory  which could then provide information about how to write, e.g., an action also in  $AdS_3$, Schroedinger 
and  the semi-flat Schroedinger geometries  discussed in  \cite{Nilsson:2013fya, Nilsson:2015x}. If this turns out to work it would give additional support for a "sequential $AdS/CFT$" 
phenomenon as suggested in  \cite{Nilsson:2012ky} where Neumann boundary conditions and the associated dynamical conformal boundary theories play a crucial role.

\acknowledgments

I am grateful to Misha Vasiliev for discussions and comments on  the manuscript. I  also thank Hampus Linander for discussions.

%%%%%%%%%%%%%%%%%%%%%%%%%%%%%%%%%%%%%%%%%%%%%%%%%%%%%%%%%%%%

%%%%%%%%%%%%%%%%%%%%%%%%%%%%%%%%%%%%%%%%%%%%%%%%%%%%%%%%%%%%

%\appendix

%%%%%%%%%%%%%%%%%%%%%%%%%%%%%%%%%%%%%%%%%%%%%%%%%%%%%%%%%%%%

%\section{Appendix}

%%%%%%%%%%%%%%%%%%%%%%%%%%%%%%%%%%%%%%%%%%%%%%%%%%%%%%%%%%%%

\end{document}